# ARTICLE

# Crosslinks increase the elastic modulus and fracture toughness of gelatin hydrogels

Anshul Shrivastava[a], Namrata Gundiah[a,b]*

Hydrogels have the ability to undergo large deformations and yet fail like brittle materials. The development of biocompatible hydrogels with high strength and toughness is an ongoing challenge in many applications. We crosslinked bovine gelatin using glutaraldehyde (control) and methylglyoxal (MGO) and assessed changes in their fracture toughness. Swelling experiments show ~710% retention of water in MGO hydrogels as compared to ~450% in control specimens. We used FTIR to identify the presence of chemical groups that may be involved in the crosslinking of gelatin gels. Scanning electron micrographs of lyophilized MGO hydrogels show large pores with plate-like intact walls that help retain water as compared to control specimens. Monotonic compression tests demonstrate nonlinear stress-strain behaviors for both hydrogel groups. MGO samples had 96% higher moduli as compared to control hydrogels that had moduli of 4.77 ± 0.73 kPa (n=4). A first order Ogden model fit the stress-strain data well as compared to Mooney-Rivlin and neo-Hookean models. We used cavitation rheology to quantify the maximum pressure for bubble failure in the hydrogels using blunt needles with inner radii of 75, 150, 230, and 320 $\mu$m respectively. Pressures inside the bubbles increased linearly with time and dropped sharply following a critical value. Bubbles in MGO gels were small and penny-shaped as compared to large spherical bubbles in control samples. We used the critical pressures to quantify the fracture energies of the hydrogels. MGO treatment increased the fracture energy by 187% from 13.09 J/m$^2$ for control gels. Finally, we discuss the challenges in using the Ogden and Mooney-Rivlin models to compute the failure energy for gelatin hydrogels.

## Introduction

Gelatin, comprised of denatured collagen, forms a three dimensional and cross-linked network when swollen with water, and undergoes large nonlinear deformations under compressive loads. Sparsely crosslinked biopolymer networks primarily bear mechanical loads in the hydrogels and transmit forces over large distances. Hydrogels of gelatin have ordered triple-helical segments, stabilized by intermolecular hydrogen bonds, and interconnected regions of flexible proteins that help the network undergo large deformations during uncoiling[1]. Entropic force generating mechanisms in hydrogels are generally described using rubber-like constitutive models[2,3]. Poroelastic and viscoelastic models have also been used to characterize the properties of water-swollen polymer hydrogels when used in drug elution and tissue engineering applications[4,5]. Hydrogels of gelatin are widely used in many applications due to their biodegradability and biocompatibility, and the ubiquitous presence of collagen in many animal tissues[6,7]. Such hydrogels however have low elastic moduli, and remarkably low fracture toughness (~10 J/m$^2$) as compared to natural hydrogels, such as cartilage and cornea, that have significantly higher values (~1000 J/m$^2$)[8–10]. Low toughness and brittle failures in gelatin hydrogels limit their applicability as scaffold materials in tissue engineering that warrant the fabrication of tougher materials, and a better understanding of the failure mechanisms in crosslinked biopolymers[6]. Porosities in the biopolymer hydrogel networks contribute to brittle failures[11].

Covalent enzymatic cross-links in the lysine/ hydroxylysine regions of the triple helical structure of collagen occur *in vivo* during tissue development in the presence of lysyl oxidase[12]. Stable trivalent crosslinks assist in the molecular slip and stretch of collagen networks and permit tissues withstand mechanical loads. Non-enzymatic advanced glycation endproducts (AGE), resulting from the oxidation of collagen with glucose, increase during aging and other pathological conditions. Metabolic by-products of the glycation reaction, including methyl glyoxal (MGO) and 3-deoxyglucosone, are associated with a higher collagen fluorescence due to the formation of intermolecular crosslinks that alter tissue insolubility and the biomechanical properties of collagenous tissues[12,13]. Tissues with higher AGE's have increased packing density of collagen fibrils, greater fibril diameters, and a consequently higher tensile strength[13]. Recent approaches to translate features at the molecular-scale, such as chemical cross-links and physical topological entanglements, to control the bulk properties of soft materials are promising in the mechanics of soft matter[14–17].

This work aims to characterize the role of crosslinks in the fracture toughness of gelatin hydrogels. Crosslinking of gelatin with glutaraldehyde is essential to produce insoluble hydrogels with mechanical integrity to withstand large deformations. Large amounts of glutaraldehyde however result in cytotoxicity and cell death[18,19]. We hence use MGO to introduce additional covalent crosslinks in glutaraldehyde treated gelatin hydrogels. We compare differences in the deformation behaviors of the crosslinked hydrogels using monotonic compression tests, and quantify their failure responses.

[a.] Department of Mechanical Engineering, Indian Institute of Science, Bangalore, India.
[b.] Centre for Biosystems Science and Engineering, Indian Institute of Science, Bangalore, India.
*For correspondence: Namrata Gundiah, namrata@iisc.ac.in, ngundiah@gmail.com

We use cavitation rheology, developed for elastomers and hydrogels, which uses the critical pressure within a bubble prior to unstable growth, to quantify the material failure[20,21]. This value of pressure depends on the elastic modulus, the initial defect size, and is related to the critical energy release rate ($G_c$) for fracture. Our results show that gelatin hydrogels crosslinked with MGO have significantly higher elastic moduli and fracture toughness as compared to control hydrogels. We interpret these differences based on changes in the swelling and microstructure of the gels due to MGO treatment and demonstrate the role of increased crosslinking in the failure of gelatin hydrogels.

## Materials and Methods

### Hydrogel fabrication

A clear solution of gelatin was prepared by dissolving bovine gelatin powder (Type-B; Sigma Aldrich, India, Bloom No. 225) in distilled water (0.05 gm/ml) at 50° C. Gelatin is soluble in water and does not undergo gelation at room temperature. We added glutaraldehyde (1% v/v of 25% stock; Sigma Aldrich, India) to stabilize the gelatin structure at room temperature. The uncured mixture was poured into a cylindrical mold (16 mm diameter X 10 mm height) and the samples were cured at 4° C for 6 hours. Specimens were soaked in 0.1 M glycine solution to block the action of glutaraldehyde and stored in distilled water overnight[2]. These samples are referred to as control hydrogels in this study.

MGO crosslinks collagen fibrils through the formation of non-enzymatic intermolecular crosslinks, and is a natural by-product during aging[12,22]. Control gelatin hydrogels, prepared earlier, were incubated in 20 mM Methylglyoxal (MGO; Sigma-Aldrich, M0252) solution at 36°C for 4.5 hours in 100 mM EPPS solution (4-(2-hydroxy-ethyl)-1-piperazinepropanesulfonic acid; Sigma-Aldrich, E1894) in phosphate-buffered saline (PBS), and titrated to a pH of 8.5[23]. Specimens were washed in PBS to block the crosslinking reaction and stored overnight in buffer at 4° C for the mechanical experiments. Fig. 1a shows representative samples from control and MGO treated groups that are visibly different.

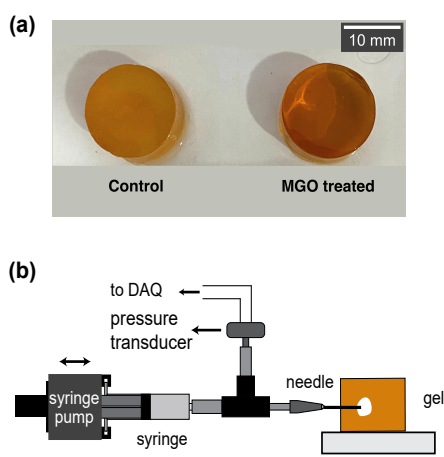

**Figure 1 (a)**: Control and MGO treated gelatin hydrogel samples show visual differences. **(b)** The cavitation setup includes a syringe attached to a syringe pump that is displaced at a controlled rate. A blunt-tipped needle is used to inflate a bubble in the hydrogel specimen. Dynamical changes in the pressure in the bubble are monitored using a pressure sensor connected to a data acquisition system.

### Swelling of hydrogels

Lyophilized control and MGO gelatin hydrogel specimens (n=3 each group) were weighed ($W_d$) and hydrated in 15 ml distilled water at room temperature[2]. The wet specimen weights ($W_s$) were measured periodically after gently patting the samples dry. Swelling ratios were calculated as

$$Swelling\ ratio = \frac{W_s - W_d}{W_d} \times 100\ \% \qquad (1)$$

The equilibrium water content (EWC) of the hydrogels was calculated using specimen weight in the maximum swollen state, $W_e$, as

$$EWC = \frac{W_e - W_d}{W_e} \times 100\ \% \qquad (2)$$

### Uniaxial compression tests

Monotonic compression tests were performed using a Bose Electroforce® 3200 (Bose Corp., USA) testing instrument equipped with two parallel compression platens (n=4 in each group). Hydrogel specimens were preloaded to 5 grams, preconditioned for 30 cycles of 10% compressive strain at 0.05 Hz, and compressed at 0.01 s$^{-1}$ until failure based on protocols reported earlier for the testing of gelatin hydrogels[2]. Loads were measured using a 25 N load cell (Honeywell Sensotec Inc., Columbus, OH, USA). Engineering stresses were calculated using the initial cross-sectional area of the hydrogel. Engineering strains were calculated based on changes in the sample heights during loading. The elastic moduli were obtained using the linear region (1-5%) of the experimentally obtained stress-strain curves. These data were fit to rubber constitutive models, including the neo-Hookean, Mooney-Rivlin, and a first order Ogden model, in MATLAB (R2020b, Natick, MA) using the function *lsqnonlin*. The goodness of fits were determined using $r^2$ values. Elastic moduli, calculated using fits to the stress-strain curves, were compared with experimentally obtained values using an unpaired t-test for all specimens in the control and MGO treated hydrogels. p values were used to determine the level of significance in these comparisons. *** denotes p-value < 0.001, ** for p < 0.01, and * p < 0.05 in this study.

### Cavitation rheology of gelatin hydrogels

Cavitation rheology was performed to quantify failures in the gelatin hydrogels. This method involves the creation of a bubble in the hydrogel using a blunt tipped needle. The maximum pressure in the bubble during inflation was measured and used to compute the fracture toughness of the gels[24] (Fig. 1b). Needles of different gauge sizes (CML Supply, Lexington, KY 40505 USA), including 75 μm, 150 μm, 230 μm, and 320 μm respectively, were inserted in the hydrogels. To prevent air leakage along the walls of the needle during bubble inflation, a small amount of gelatin mixed with 1% glutaraldehyde was poured along the needle walls prior to insertion. This procedure allowed formation of air bubbles ahead of the needle tip alone as compare to leakage of air along the needle length. The needle was connected to a 20 cc syringe, fixed to a syringe pump (NE1000, New Era Pump Systems). The bubble was inflated at 2000 μL/min until failure. The real-time pressure within the bubble was recorded using a pressure sensor (ASDXAVX030PGAA5, Honeywell Sensotec Inc., Columbus, OH, USA) and an Arduino UNO board. Temporal variations in pressure within the bubble were obtained, and the critical pressure, $P_c$, was identified as the maximum pressure in the bubble prior to collapse.

## Scanning Electron Microscopy

Hydrogel surfaces were imaged using a Scanning Electron Microscopy (SEM) to explore microstructural differences due to the crosslinking methods in this study. Pure gelatin, control (treated with 1% glutaraldehyde), and MGO samples were prepared as described earlier, lyophilized, mounted on an aluminium stub using double sided carbon tape, and sputter coated with a thin layer of gold. A scanning electron microscope (ΣIGMA™ FESEM, Carl Zeiss NTS Ltd, Cambridge, UK), with accelerating voltages from 5 to 10 kV was used to image the specimen surfaces.

## FTIR spectroscopy

Samples of pure gelatin, controls (treated with 1% glutaraldehyde), and MGO crosslinked specimens were prepared as described earlier. Test samples were lyophilized, ground into a powder, and placed in the spectrophotometer. A single beam Fourier Transform Infrared (FTIR) spectrophotometer (Alpha II, Bruker Scientific LLC, USA) was used in transmittance mode (4000-600 cm$^{-1}$) to characterize the vibrational spectra of the hydrogel samples with a scanning resolution of 4 cm$^{-1}$. Spectra were performed in a dry atmosphere at room temperature and reported after subtracting from the background.

## Visualization of bubble inflation

Videography of bubble inflation in the gelatin gels was performed at 2000 fps using a high-speed color camera (Phantom VEO-E 310L). The camera was mounted above the sample, illuminated from the base using an LED light source, and the time taken for growth of bubbles after inflation was measured to visualize the dynamics of bubble expansion until failure in control and MGO treated gels.

## Results and Discussion

### MGO treatment alters the swelling properties and nonlinear mechanical behaviors of gelatin hydrogels

Fig. 2 shows the swelling behaviors of control and MGO treated gelatin hydrogels over a 23-hour period in distilled water (n=3 each group). Pure gelatin specimens were fragile to handle, liquid like at room temperature, and dissolved easily in water. These specimens were hence not included in the studies. Crosslinks are essential to keep gelatin in gelated form at room temperature.

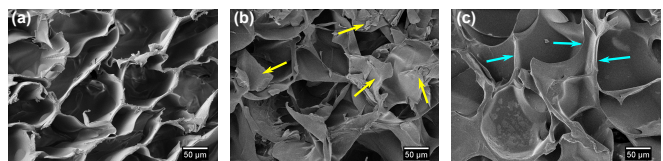

**Figure 3**: SEM images for (a) pure Gelatin, (b) control and (c) MGO treated gels show clear differences. Pure gelatin is very porous and dissolves in water. Glutaraldehyde crosslinked hydrogels (control) have pores that appear collapsed (shown by yellow arrows). In comparison, MGO sample have plate like microstructures with intact pore walls (shown by cyan arrows).

Control gelatin hydrogels had lower EWC (~450%) as compared to MGO specimens which swelled to ~ 710% at equilibrium. Dry specimen weights were higher for the MGO group by 17% as compared to the control samples (p<0.01). SEM micrographs show high number of well delineated pores for pure uncrosslinked gelatin as compared to control and MGO samples (Fig. 3). The collapsed biopolymer walls with plate-like wall architecture in control samples may be a result of the lyophilization process. Fig. 3c shows a representative MGO specimen that has plate-like features with intact walls, and the presence of large pores that assist retain water in the structure. Network porosity appears clearly altered for MGO samples as compared to pure gelatin specimens. These results agree with the swelling behaviors of the gels that demonstrate significantly higher EWC for MGO samples (Fig. 2).

We used FTIR spectroscopy to investigate changes in the hydrogel composition due to MGO treatment. Fig. 4 shows transmittance curves for pure gelatin, control, and MGO treated samples. An increase in the crosslinking density of the gelatin hydrogels using glutaraldehyde or MGO results in a proportional shift in the curves from pure gelatin samples. The main peaks labeled in the curves correspond to O-H bond stretching (~ 3305 cm$^{-1}$), C-H bond stretching (~2962 cm$^{-1}$), secondary amide bond (~ 1640 cm$^{-1}$), and C-H bond bending (~ 1451 cm$^{-1}$). A decrease in the transmittance peaks corresponding to these bonds suggests their involvement in crosslinking gelatin with glutaraldehyde and MGO. The proportional changes in the amplitudes corresponding these bonds also demonstrate greater crosslink densities for MGO treated samples as compared to control, and pure gelatin samples.

Nonlinear mechanical tests may be obtained using monotonic tensile or compression tests[1,7,25,26]. Because hydrogel samples are highly compliant and fragile when hydrated, this poses experimental

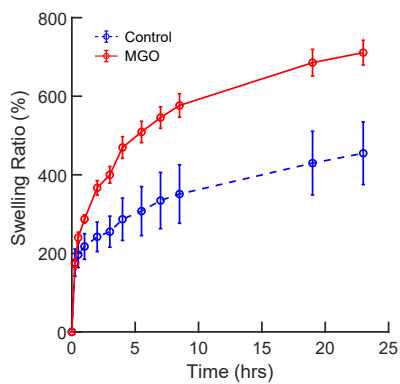

**Figure 2**: Swelling behaviors of the control and MGO gels in distilled water. MGO treated gels swelled significantly more (~710%) as compared to control gels (~450%).

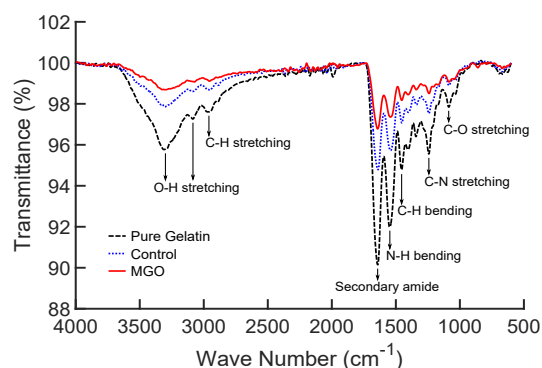

**Figure 4**: FT-IR spectra show proportional changes in the transmission results due to crosslinking by glutaraldehyde and MGO as compared to pure gelatin.

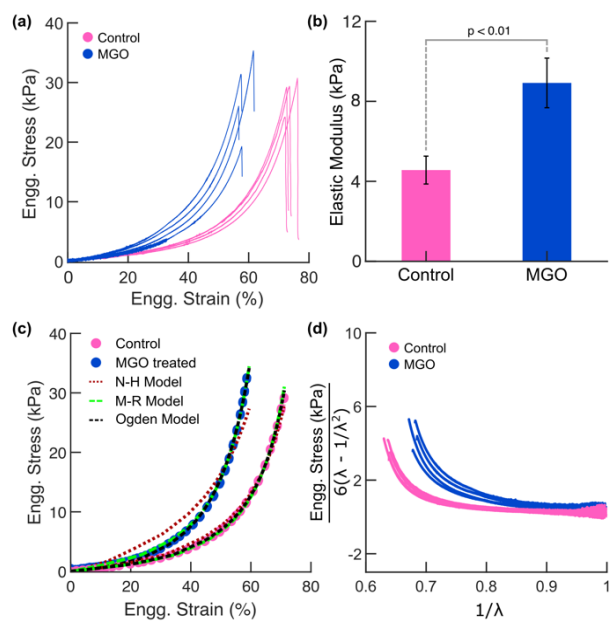

**Table 2**: Coefficients obtained from fits to the neo-Hookean, Mooney-Rivlin, and first order Ogden hyperelastic models are included below for control and MGO hydrogels.

|  |  | Control hydrogels | MGO hydrogels |
|---|---|---|---|
| neo-Hookean model | C | 0.30 ± 0.04 | 0.67 ± 0.13 |
|  | $r^2$ | 0.99 ± 0.002 | 0.957 ± 0.01 |
|  | $E_{NH}$ | 1.82 ± 0.21 | 4.03 ± 0.79 |
| Mooney-Rivlin model | $C_{01}$ | 0.19 ± 0.03 | - 0.01 ± 0.08 |
|  | $C_{10}$ | 0.04 ± 0.01 | 0.34 ± 0.08 |
|  | $r^2$ | 0.99 ± 0.001 | 0.99 ± 0.003 |
|  | $E_{MR}$ | 1.38 ± 0.19 | 1.99 ± 0.44 |
| First order Ogden model | $\mu_p$ | 1.07 ± 0.16 | 1.22 ± 0.23 |
|  | $\alpha_p$ | 2.92 ± 0.16 | 5.02 ± 0.25 |
|  | $r^2$ | 0.99 ± 0.000 | 0.99 ± 0.000 |
|  | $E_{OM}$ | 4.69 ± 0.61 | 9.18 ± 1.81 |

**Figure 5 (a)**: Engineering stress- engineering strain data from compression tests (n=4 each group) show nonlinear and large deformation behaviors. **(b)** Elastic moduli, calculated using the initial linear region of the stress-strain curves, were significantly higher for MGO treated samples as compared to controls (p<0.01). **(c)** We used a neo-Hookean, Mooney-Rivlin, and first order Ogden nonlinear constitutive models to fit the experimental data. **(d)** Compression data were used to assess the suitability of the Mooney-Rivlin model for samples from control and MGO groups in the study.

challenges during clamping specimens for tensile tests. We hence used monotonic uniaxial compression tests to compare changes in the material properties of the hydrogels. All gelatin hydrogels fragmented into small irregular shaped pieces as reported earlier for other hydrogel specimens[27]. Mechanical tests show clear differences in the non-linear mechanical responses of control and MGO gelatin hydrogels (Fig. 5a). Engineering stress – engineering strain curves show clearly stiffer response for the MGO group as compared to the control hydrogels (Fig 5a, n=4 for each group). All curves were J-shaped and include an initial toe region accompanied by a region of large deformation, and a stiffening response. A sharp drop in stress seen in all test results shows the point of brittle gel failure. The maximum strain to failure and elastic modulus, obtained as the slope of the linear region of stress-strain curve, were quantified using the experimental data (Table 1). Cross-linking of the triple helical structures in gelatin restrict movement of the protein molecules which influence the material modulus and failure properties[1,2].

**Table 1**: Results from uniaxial compression tests are shown from control and MGO hydrogels.

| Control hydrogels | | | | | |
|---|---|---|---|---|---|
| Sample | 1 | 2 | 3 | 4 | Mean ± Std |
| Max. Strain (%) | 73.5 | 72.03 | 72.61 | 76.16 | 73.39 ± 1.63 |
| Max. Stress (%) | 29.4 | 24.2 | 29.2 | 30.7 | 27.3 ± 3.5 |
| $E_{exp}$ (kPa) | 3.85 | 4.83 | 4.8 | 5.64 | 4.77 ± 0.73 |
| MGO hydrogels | | | | | |
| Max. Strain (%) | 61.57 | 57.69 | 56.55 | 57.5 | 58.42 ± 1.93 |
| Max. Stress (%) | 35.3 | 19.3 | 26.1 | 31.4 | 26.84 ± 6.58 |
| $E_{exp}$ (kPa) | 10.27 | 7.43 | 10.05 | 9.62 | 9.34 ± 1.30 |

The maximum strain to failure was significantly lower for MGO treated gels (~ 58%) as compared to control gels (~73%). MGO treated gels also had significantly higher (p<0.01) elastic moduli (9.34 ± 1.30 kPa) as compared to control gels (4.77 ±0.73 kPa) (Fig. 5b). There were no statistical differences in the maximum stress between the two groups due to MGO treatment.

We fit the non-linear compression response of the hydrogel samples using neo-Hookean, Mooney-Rivlin and first order Ogden models (Table 2). Fig 5c shows fits for one representative sample in the MGO and control groups respectively. The goodness of fits to the different models were based on the $r^2$ values. These results show that the Mooney-Rivlin and first order Ogden models provided good fits to the data from control and MGO treated gels ($r^2$~ 0.9995). $r^2$ values were lower for MGO samples fitted using the neo-Hookean model (0.9573) as compared to those in the control group. Uniaxial compression data from control and MGO groups were used to check for the validity of the Mooney–Rivlin equation. Engineering stresses, $\sigma_{11}$, computed for the Mooney-Rivlin model are expressed in terms of the compressive stretch, $\lambda_1$, as

$$\sigma_{11} = 6\left(\lambda_1^2 - \frac{1}{\lambda_1}\right)\left(C_{01} + \frac{C_{10}}{\lambda_1}\right) \quad (3)$$

$C_{01}$ and $C_{10}$ are constants in the above equation. Fig. 5d shows plots of $\sigma_{11}/6\left(\lambda_1^2 - \frac{1}{\lambda_1}\right)$ with $\frac{1}{\lambda_1}$ using the experimental data for the control and MGO gelatin hydrogels. $C_{10}$, obtained using the slope of the plot, was positive for both control and MGO samples (Table 2). In contrast, $C_{01}$ was negative for MGO samples. The adscitious Baker-Ericksen inequalities for an isotropic homogeneous material, with strain energy function described using the Mooney-Rivlin model, are given by $C_{01}, C_{10} \geq 0$. These inequalities ensure that the direction of greater stress occur in the direction of loading where the stretch values are high[28,29]. The elastic modulus computed using Mooney-Rivlin model was also ~78% lower for MGO, and ~70% lower for control samples as compared to experimental data. We did not hence consider the Mooney-Rivlin model in the remaining part of this study.

We next compared the elastic moduli computed from the neo-Hookean model with experimentally obtained values using the linear region of the stress-strain data. The moduli calculated for the gelatin hydrogels using the neo-Hookean models were significantly lower for the MGO and control samples as compared to experimental results (Table 2). In contrast, there were no differences between experimental results for elastic moduli from those calculated using the Ogden model. These results show that the Ogden model fit the mechanical properties of the hydrogels better than the neo-Hookean one which is generally reported in the literature for hydrogels[30,31].

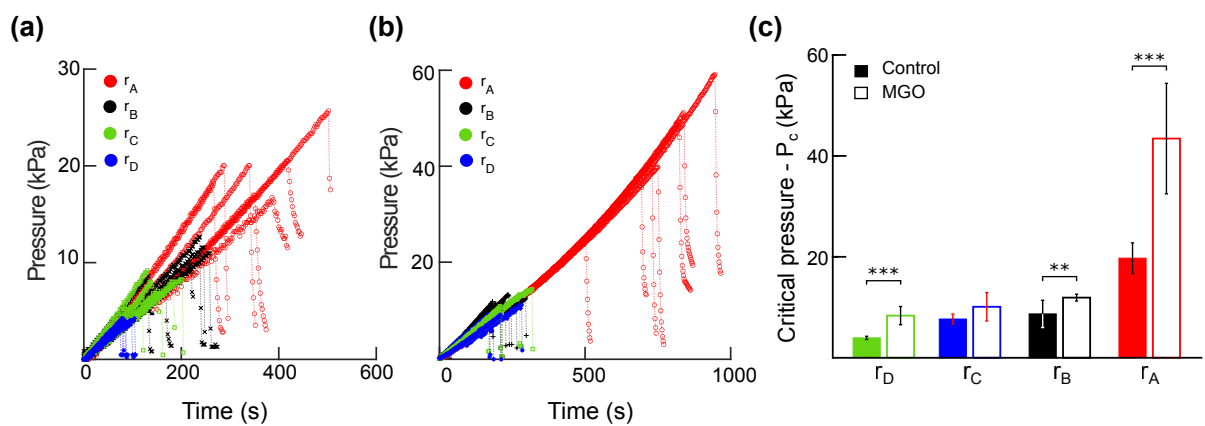

**Figure 6**: Pressure variations within the bubbles are plotted with inflation time for **(a)** control and **(b)** MGO treated gels using needles with inner radii corresponding to $r_A$ (75 μm), $r_B$ (150 μm), $r_C$ (230 μm), and $r_D$ (320 μm). **(c)** The critical pressure, obtained as the maximum pressure during inflation, was compared during cavitation for control and MGO treated gelatin hydrogel samples for each of the different needles in the study (n=7 each needle group).

### Critical pressure for bubble failure using cavitation rheology

To assess the fracture properties of gels, blunt needles of various radii ($r_A$: 75 μm, $r_B$: 150 μm, $r_C$: 230 μm, and $r_D$: 320 μm) were inserted in the gels to create flaws of known sizes as described earlier. The pressure variations during bubble inflation are shown for control (Fig. 6a) and MGO (Fig. 6b) treated gels for each of the different needle sizes in the study (n=7 for each group). Pressures inside the bubble increased linearly with time for most hydrogels until the point of unstable growth which resulted in failure. The time taken for bubble collapse was significantly higher for each of the different needles in MGO treated gels as compared to those in the control group. Interestingly, the results from the smallest needle ($r_A$) for MGO samples showed nonlinear pressure increase during inflation. Results were also more repeatable for MGO samples as compared to control hydrogels. We obtained the critical pressures, $P_c$, based on the maximum value prior to failure, and compared these values for control and MGO hydrogels. The values of $P_c$ were significantly higher for the smallest needle diameters for both hydrogel groups. $P_c$ was significantly higher (p<0.01) for MGO gels as compared to control samples for all needle diameters (Fig. 6c) barring those corresponding to $r_C$ that showed higher variance. These results correlated with the significantly higher compression moduli for MGO specimens as compared to control hydrogels (Table 1).

Cavity expansion is an attractive technique to study the local mechanical response of elastomers and hydrogels. Gent and Lindley first demonstrated the importance of cavitation based on observations of internal flaws in vulcanised rubbers under tensile loads[32]. Cavitation generally occurred near a spherical inclusion in elastomers[31]. Gas bubbles in elastomers generally form under pressure due to the release of entrapped gas during preparation[34,35]. Analytical expressions for fractures in elastomers are based on an energy criterion using the condition for propagation of a pressurised crack in the sample[36]. Zimberlin and co-workers used needle induced cavitation to calculate the local modulus of a hydrogel based on the pressure required for unstable gel failure[37]. This method, called cavitation rheology, was later used to develop scaling relationships and characterize the fracture behaviors of polyacrylamide hydrogels[34]. More recently, cavitation rheology has been used to measure the spatially localised mechanical properties of biological tissues, and to simulate the damage mechanisms in traumatic brain injuries[20,38,39].

### Crosslinks contribute to the higher toughness of MGO treated hydrogels

We used the values of the critical pressure, $P_c$, for hydrogel failure to compute the fracture energies for the two hydrogel groups in our study. Fractures occur when the applied energy release rate during bubble inflation is equal to the critical energy release rate ($G_c$). The critical pressure for cavitation ($P_c$) is related to the critical energy release rate for fracture ($G_c$) as [24,40]

$$\frac{P_c}{\sqrt{E}} = \sqrt{\left(\frac{\pi G_c}{3}\right)} \sqrt{\left(\frac{1}{r}\right)} \quad (4)$$

The equation 4 was obtained for a linear elastic material, and is used to study failures due to cavitation in elastomers and hydrogels[24,40].

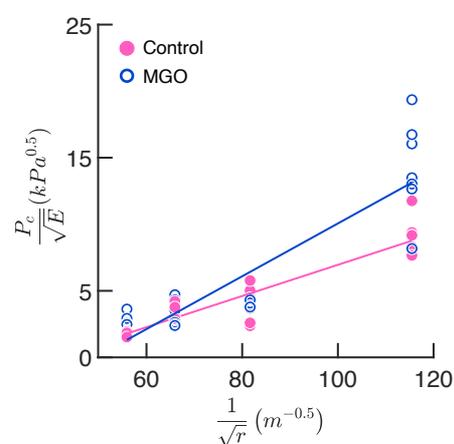

**Figure 7**: Variation of critical pressure with the inverse of needle inner radius is shown for MGO-treated and control gels. Critical energy release rate ($G_c$) is given by the slope of the $P_c/\sqrt{E}$ v/s $1/\sqrt{r}$ plot.

We used the experimentally obtained elastic moduli, E, for both hydrogel groups and plotted the variations in $P_c/\sqrt{E}$ with $1/\sqrt{r}$ for each of the different hydrogels in the study. The slope of the curves in Fig. 7 had $r^2=0.96$ for control and $r^2=0.89$ for MGO treated hydrogels. The experimentally obtained critical energy release rate ($G_c$), given by the slope, was 13.09 J/m$^2$ and 37.59 J/m$^2$ for the control and MGO hydrogels, respectively. The higher variability in critical pressures obtained in our study suggests the importance of the local microstructure in the failure process. Although a localised measurement of the failure parameters is useful to assess the toughness of hydrogels, a single value of toughness does not capture possible variations in the microstructure. Heterogeneity in the molecular weights of commercially available gelatin powder from different sources may introduce variability in the measured properties of gelatin. Mechanical quantification of the failure processes in gelatin hydrogels using cavitation rheology hence demand multiple experiments

Cavitation is a complex phenomenon and requires careful experimental methods for needle insertion and formation of bubbles ahead of the needle, and suitable analytical formulation to quantify hydrogel failures. We used high speed videography to assess the dynamics of bubble inflation during cavitation rheology (Movie 1, Movie 2). Bubble dynamics for cavitation include a phase for initiation, and a second one related to growth. Fig. 8 shows changes in the bubble growth, following bubble initiation, for control and MGO treated hydrogels. These results show clear differences in the size and shape of the bubbles in the control specimen as compared to MGO hydrogel. Bubbles in the control hydrogels generally grow symmetrically and are spherical as compared to penny shaped bubbles in the MGO treated samples. The time required for failure following initiation was 317.2 ± 49.8 ms in the control hydrogels which was significantly higher (p<0.05; n=3) than the corresponding time measured for the MGO hydrogels (168.3 ± 29.3 ms). Although the total time for failure was higher for the MGO gels compared to controls (Fig. 6a and 6b), the time for failure after bubble initiation was significantly shorter for MGO samples. These results show the role of crosslinking in the failure characteristics of gelatin hydrogels.

Different techniques, such as laser induced cavitation and acoustic induced cavitation, have been used in literature to induce cavitation in elastomers and hydrogels in addition to needle induced cavitation[41–43]. The analytical expression used in cavitation assumes the growth of spherical bubbles in an incompressible, isotropic and hyperelastic material which is described using a strain energy density function, $W(\lambda_\theta)$, where $\lambda_\theta$ is the circumferential stretch[35]. The fracture toughness measurement by cavitation is based on the external work of pressurization which is equated to the change in elastic energy, and surface energy due to the creation of new interfaces during bubble inflation[24,35,40,44]. The elastic energy term due to bubble inflation during cavitation is obtained assuming a specific form of $W(\lambda_\theta)$ given by[31,45,46]:

$$P = \int_1^\lambda \frac{1}{(\lambda_\theta^3-1)} \frac{dW(\lambda_\theta)}{d\lambda_\theta} d\lambda_\theta \qquad (5)$$

For a neo-Hookean material, with elastic modulus, E, equation 5 simplifies to[30,45]

$$P = E\left[\frac{5}{6} - \frac{2}{3\lambda_\theta} - \frac{1}{6\lambda_\theta^4}\right] \qquad (6)$$

Earlier studies show that the elastic instability during cavitation due to an initial flaw of radius, r, is associated with a linear increase in $P_c/E$ values with $\gamma/Er$ given by[34,37]

$$\frac{P_c}{E} = \frac{5}{6} + \frac{2\gamma}{Er} \qquad (7)$$

Contribution from the surface energy term is neglected for a material with high elastic modulus or when a large needle is used to induce cavitation bubbles. In contrast, for a material with low elastic modulus, or for cases when small needle radii are used, the surface energy term is of the same magnitude as the flaw size. A significantly higher pressure is hence required for cavity growth and rupture[22].

Because the first order Ogden model fit the experimentally obtained stress-strain results for control and MGO samples better as compared to other models in our study (Fig. 5c; Table 2), we investigated changes to the analytical expressions for pressure in the fracture energy calculation during cavitation using this material model. The strain energy density function for a first order Ogden material is given by (Appendix):

$$W(\lambda_\theta) = \frac{\mu_p}{\alpha_p}\left(\lambda_\theta^{-2\alpha_p} + 2\lambda_\theta^{\alpha_p} - 3\right) \qquad (8)$$

where $\mu_p$ and $\alpha_p$ are the model constants for the control and MGO hydrogels (Table 2). The variation in pressure in the bubble for the control gels is given in terms of $\lambda_\theta$ as (Appendix):

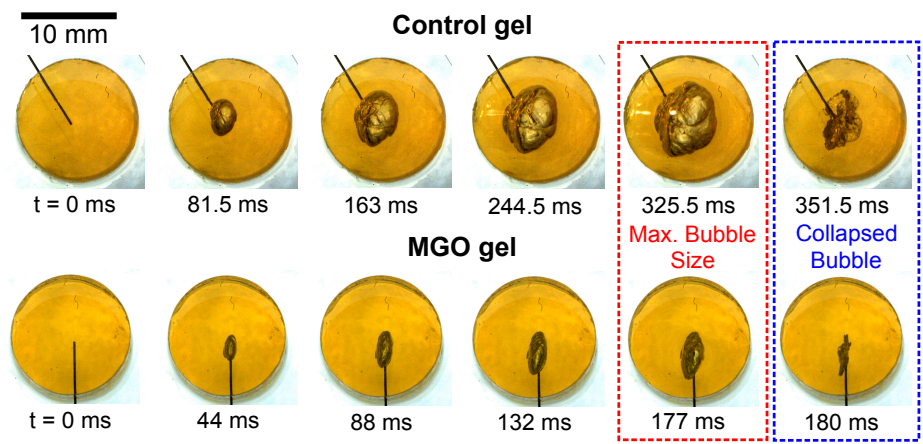

**Figure 8**: High speed videography from control and MGO gels are shown from bubble initiation time until failure. Rough surfaces are evident during inflation in the hydrogels. The growth of the bubble is dramatically different in the control and MGO hydrogels.

$$P_{Control} = 2.046 \left[ ln(\lambda_\theta) - \frac{1}{6\lambda_\theta^6} - \frac{1}{3\lambda_\theta^3} + \frac{1}{2} \right] \quad (9)$$

The corresponding values for MGO treated hydrogels are given by

$$P_{MGO} = 2.446 \left[ \frac{139}{140} + \frac{\lambda_\theta^2}{2} - \frac{1}{4\lambda_\theta^4} - \frac{1}{7\lambda_\theta^7} - \frac{1}{10\lambda_\theta^{10}} - \frac{1}{\lambda_\theta} \right] \quad (10)$$

The expressions for critical pressure for cavitation become unbound as $\lambda_\theta \to \infty$ in contrast to the bounded solutions that were obtained when assuming a neo-Hookean material model. We note a similar problem when using the Mooney-Rivlin model to obtain the expression for critical pressure in gelatin hydrogels. The analytical approach for cavitation using an Ogden model cannot hence be used to compute the fracture toughness of hydrogels which poses a serious limitation in cavitation rheology.

Numerous methods have been used to develop tougher and stiffer hydrogels[15,48–52]. Gong and co-workers developed interpenetrating double network hydrogels (DN hydrogels) that undergo large deformations. Inter-penetrating hydrogels combine the properties of crosslinked polymers in providing sacrificial bonds under deformation for energy dissipation with a second network that can withstand large deformations[14,15]. Energy dissipation during the breakage of sacrificial bonds results in soft gels with high failure energies. A combination of ionic and covalent crosslinks is also used to toughen hydrogels[14]. Crosslinking of gelatin gels using high concentrations of glutaraldehyde is cytotoxic to cells[18,19]. In contrast, MGO is one of the metabolites produced *in vivo* which reduces adverse immunoresponse in cells[22,53]. Snedeker and co-workers used MGO to crosslink collagen fibrils in rat tail tendons[23]. These studies showed a higher yield stress and reduced stress relaxation in the MGO treated tissues. Our studies also show increased fracture toughness of gelatin hydrogels crosslinked using MGO. Svenssen *et al* used sodium borohydrate to crosslink gelatin[54]. This method produces hydrogen as a byproduct in the homogeneous gelatin hydrogels. Because macroscopic pores influence the fracture properties, we did not use this method to crosslink hydrogel samples in this study. Results from our study hence demonstrate that MGO treatment is useful to produce gelatin hydrogels with high modulus and high fracture toughness which may be useful in the development of soft actuators, for tissue engineering and drug delivery, and in mechanobiological studies on substrates of tunable stiffness.

## Conclusions

Crosslinking of gelatin hydrogels using MGO presents many advantages due to reduced cytotoxicity and biocompatibility. (i) FT-IR spectroscopy shows a shift of the transmittance spectra corresponding to covalent bonds in gelatin that may be involved in crosslinking of MGO gels. The higher density of non-enzymatic crosslinks in these samples increased the experimentally measured material modulus. MGO treated hydrogels have higher (96%) elastic modulus as compared to control gels that have moduli of 4.77±0.63 kPa. The Ogden model fit the experimental results from control and MGO gelatin samples well. (ii) MGO treated gels absorb significantly higher amounts of water as compared to the glutaraldehyde crosslinked control gels which makes it an attractive carrier for drug eluting applications. SEM micrographs show clear differences in pure gelatin and glutaraldehyde treated samples as compared to MGO specimens. Plate like microstructure in MGO hydrogels correlated with increased mechanical strength, whereas the larger pores facilitate higher water absorption. (iii) Gel mechanical behaviors were more repeatable for the MGO treatment as compared to control samples. Cavitation studies show that all gels failed catastrophically on reaching a critical pressure which is significantly higher for MGO treated gels compared to control hydrogels. Fracture toughness of the hydrogels, obtained using a linear elastic material model with experimentally determined critical pressures, were significantly higher (~165%) for the MGO samples as compared to control hydrogels. Analytical models for the quantification of hydrogel toughness using cavitation rheology that are based on the Ogden material model show that the critical pressures are unbound. Crosslinking of gelatin gels with glutaraldehyde and MGO is an attractive method to produce hydrogels with higher moduli and increased toughness.

## Author Contributions

AS performed experiments, analyzed results, and helped write the manuscript. NG designed the study, analyzed the results, and wrote the manuscript.

## Conflicts of interest

There are no conflicts to declare.

## Acknowledgments

NG is grateful to the Department of Biotechnology (BBI2) for project support. Thanks to Akshay Sharma, Venkitesh and Dr. Susmita Dash for help with high-speed videography, and to Achu Byju for preliminary help with the cavitation experimental setup. We also thank Gulista Bano for help with FT-IR experiments. AS is supported through a PMRF fellowship from MHRD, Government of India.

## Appendix

**Calculation of variation of pressure with circumferential stretch for Ogden model**

We assume spherical and symmetric expansion of a bubble within a hyperelastic, isotropic, homogeneous, and incompressible material. The inner surface of the bubble is subjected to an equi-biaxial deformation. The principal stretches in spherical polar coordinates have the relation

$$\lambda_r \lambda_\theta \lambda_\phi = 1 \quad (A1)$$

$\lambda_r$ is the principal stretch in the radial direction. $\lambda_\theta$, the circumferential stretch, is equal to $\lambda_\phi$ due to spherical symmetry. Thus,

$$\lambda_r = \frac{1}{\lambda_\theta^2} \quad (A2)$$

For a first order Ogden model, we write

$$W(\lambda_r, \lambda_\theta, \lambda_\phi) = \frac{\mu_p}{\alpha_p} \left( \lambda_r^{\alpha_p} + \lambda_\theta^{\alpha_p} + \lambda_\phi^{\alpha_p} - 3 \right) \quad (A3)$$

Using eqn. (A2) in eqn. (A3), we write

$$W(\lambda_\theta) = \frac{\mu_p}{\alpha_p} \left( \lambda_\theta^{-2\alpha_p} + 2\lambda_\theta^{\alpha_p} - 3 \right) \quad (A4)$$

Differentiating this equation with respect to $\lambda_\theta$, we write

$$\frac{dW(\lambda_\theta)}{d\lambda_\theta} = 2\mu_p \left(\lambda_\theta^{\alpha_p-1} - \lambda_\theta^{-2\alpha_p-1}\right) \tag{A5}$$

Using eqn. (5), we can write the variation of pressure in the bubble as

$$P = 2\mu_p \int_1^\lambda \frac{1}{\lambda_\theta^3 - 1} \left(\lambda_\theta^{\alpha_p-1} - \lambda_\theta^{-2\alpha_p-1}\right) d\lambda_\theta \tag{A6}$$

We use the average values of $\mu_p = 1.023$ and $\alpha_p = 3$ (rounded), for control gels (Table 2) to obtain

$$P_{Control} = 2.046 \left[ ln(\lambda_\theta) - \frac{1}{6\lambda_\theta^6} - \frac{1}{3\lambda_\theta^3} + \frac{1}{2} \right] \tag{A7}$$

Similarly, using the average values of $\mu_p = 1.223$ and $\alpha_p = 5$ (rounded) for MGO gels, we have

$$P_{MGO} = 2.446 \left[ \frac{139}{140} + \frac{\lambda_\theta^2}{2} - \frac{1}{4\lambda_\theta^4} - \frac{1}{7\lambda_\theta^7} - \frac{1}{10\lambda_\theta^{10}} - \frac{1}{\lambda_\theta} \right] \tag{A8}$$